\documentclass[twocolumn, tighten]{aastex631}
\usepackage{bm}
\usepackage{amsmath} 
\usepackage{booktabs}
\usepackage{array}
\usepackage{enumerate}
\usepackage{paralist}

\renewcommand\thesection{\Roman{section}}

\newcommand{\KIAA}{\affiliation{Kavli Institute for Astronomy and
Astrophysics, Peking University, Beijing 100871, China}}
\newcommand{\DOA}{\affiliation{Department of Astronomy, School of Physics,
Peking University, Beijing 100871, China}}

\newcommand{\NAOC}{\affiliation{National Astronomical Observatories,
Chinese Academy of Sciences, Beijing 100012, China}}

\received{XXX}
\revised{YYY}
\accepted{ZZZ}
\submitjournal{ApJ}
\shorttitle{New limits on the Lorentz/CPT symmetry with gravitational waves}
\shortauthors{Z. Wang, L. Shao, and C. Liu}

\begin{document}

\title{New limits on the Lorentz/CPT symmetry through fifty gravitational-wave events}
\correspondingauthor{Lijing Shao}
\email{lshao@pku.edu.cn}
\author[0000-0002-8742-8397]{Ziming Wang}\DOA
\author[0000-0002-1334-8853]{Lijing Shao}\KIAA\NAOC
\author[0000-0001-7649-6792]{Chang Liu}\DOA\KIAA

\begin{abstract}
  Lorentz invariance plays a fundamental role in modern physics. However, tiny
  violations of the Lorentz invariance may arise in some candidate quantum
  gravity theories. Prominent signatures of the gravitational Lorentz invariance
  violation (gLIV) include anisotropy, dispersion, and birefringence in the
  dispersion relation of gravitational waves (GWs). Using a total of 50 GW
  events in the GW transient catalogs GWTC-1 and GWTC-2, we perform an analysis
  on the anisotropic birefringence phenomenon. The use of multiple events allows
  us to completely break the degeneracy among gLIV coefficients and {\it
  globally} constrain the coefficient space. Compared to previous results at
  mass dimensions 5 and 6 for the Lorentz-violating operators, we tighten the
  global limits of 34 coefficients by factors ranging from $2$ to $7$.
\end{abstract}


\keywords{General Relativity — Gravitational Waves — Compact Binary Stars}

\section{ Introduction }
\label{ sec:intro }

While the discovery of gravitational waves \citep[GWs;][]{Abbott:2016blz} is a
strong evidence in support of General Relativity (GR), it also provides new
approaches for testing GR. The classical GR is the most successful theory of
gravity to date, but it is not compatible with quantum field theory which
describes the other three fundamental interactions very well. In some candidate
theories of quantum gravity like the  string theory, the Lorentz invariance
(LI), one of the foundations of modern physics, could spontaneously break
\citep{Kostelecky:1988zi, Kostelecky:1991ak}. LI violation (LIV) could happen in
many different sectors such as the electron sector, the photon sector and the
gravity sector. Experiments bounds on these sectors can be found in
\citet{Kostelecky:2008ts}.

The standard-model extension \citep[SME;][]{Colladay:1996iz, Kostelecky:2003fs}
is a powerful and popular framework to explore LIV. SME bases on an effective
field theoretic approach and contains all possible operators for Lorentz and CPT
violations in GR and the standard model of particle physics.  In the SME, a LIV
term in the Lagrangian contains a LIV operator constructed from the contraction
of a conventional tensor operator with a violating tensor coefficient. In the
framework of SME, the symmetry breaking is spontaneous \citep{Bluhm:2004ep}, and
the violations are observer LI.

In the past years, a number of experiments have been performed to investigate
the gravitational LIV (gLIV) in SME \citep[for a review, see
e.g.,][]{Hees:2016lyw}. Their constraints come from lunar laser ranging
\citep{Battat:2007uh, Bourgoin:2017fpo}, atom interferometry
\citep{Muller:2007es}, planetary ephemerides \citep{Iorio:2012gr, Hees:2015mga},
\v{C}erenkov radiation of cosmic rays \citep{Kostelecky:2015dpa}, pulsar
timing~\citep{Shao:2014oha, Shao:2014bfa, Shao:2019nso, Jennings:2015vma,
Shao:2016ezh, Shao:2018vul, Shao:2019cyt}, short-range gravity experiments
\citep{Bailey:2014bta, Shao:2016cjk}, Very Long Baseline Interferometry
\citep{LePoncin-Lafitte:2016ocy}, GWs \citep{Kostelecky:2016kfm, 
Schreck:2016qiz, Shao:2020shv} and so on.

As shown by \citet{Kostelecky:2016kfm} and \cite{Monitor:2017mdv}, GWs provide
an opportunity for tests of LIV in the pure gravity sector.  In the SME, each
LIV term can be cataloged with a specific mass dimension $d$
\citep{Kostelecky:2009zp, Kostelecky:2018yfa}. Terms with a higher $d$ are
considered to appear at higher energy scales, which means a higher order
correction. Similar to the situation of the photon sector
\citep{Kostelecky:2002hh, Kostelecky:2009zp}, the anisotropic birefringence
phenomenon can be used to restrict the Lorentz/CPT violations tightly in
nonminimal gravity with the mass dimension of LIV operators $d \ge 5$. At the
time of \citet{Kostelecky:2016kfm}, there was only one GW event, GW150914, so at
a specific mass dimension $d$ they only had one inequality to constrain a linear
combination of coefficients. At $d =  5$ and $6$, they obtained, $\big|
\sum_{jm}Y_{jm}(\theta,\phi) \,k^{(5)}_{(V)jm} \big|\leq2\times10^{-14}\,
\text{m}$ and $\big|\sum_{jm}{_{+4}Y_{jm}} (\theta,\phi)\,\big(k^{(6)}_{(E)jm} +
ik^{(6)}_{(B)jm}\big)\big|\leq8\times10^{-9} \,\text{m}^2$,
where $\theta \approx 160^\circ$ and $\phi \approx 120^\circ$ are the rough sky
position of GW150914 in the celestial-equatorial frame; $Y_{jm}$ (also denoted
as ${_{0}Y_{jm}}$) and ${_{+4}Y_{jm}}$ are the spin-weighted spherical harmonics
that are direction dependent; $k^{(5)}_{(V)jm}$, $k^{(6)}_{(V)jm}$ and
$k^{(6)}_{(B)jm}$ are SME coefficients controlling the gLIV. The constraint at
$d = 5$ is the first constraint on gLIV operators, while the $d = 6$ constraint
is comparable to the limits from short-range laboratory tests \citep[see e.g.,
][]{Shao:2016cjk}.

Afterwards, there are 11 events published in the GW transient catalog GWTC-1
\citep{LIGOScientific:2018mvr}, which enables a global analysis to constrain all
the coefficients simultaneously. \citet{Shao:2020shv} extended the analysis
method in \citet{Kostelecky:2016kfm}  to a global approach, using the whole
GWTC-1. The analysis broke the degeneracy in SME coefficients and improved the
limits by 2--5 orders of magnitude. 

Notice that even in the mass dimension 5 there are still 16 independent
components of the violation coefficients, but 11 events could only provide 11
inequalities on the linear combination of them. Therefore, in fact these
inequalities can not fully break the parameter degeneracy at $d=5$, not to
mention the higher mass dimensions.  However, with more and more observations of
GW events, now with the addition of GWTC-2 \citep{LIGOScientific:2020ibl}, we
can \textit{completely} decouple the degenerate parameters.

In this work we will focus on constraining the gLIV coefficients using the whole
GW transient catalogs GWTC-1 and GWTC-2 \citep{LIGOScientific:2018mvr,
LIGOScientific:2020ibl}, which contains 50 GW events in total. Benefiting from
the multiple GW events coming from different sky areas, we finally break the
degeneracy among \textit{all} coefficients at $d =  5$ and $d=6$, and obtain
limits which are $2$ to $7$ times tighter than the existing best results. 

This {\it Letter} is organized as follows. In Sec.~\ref{ sec:disper } we
introduce the dispersion relation of GWs in the linearized gravity with gLIV,
and the effects of anisotropic birefringence during the propagation of GWs. In
Sec.~\ref{ sec:costrain } we select the source parameters from the posterior
samples of 50 GW events. The constraints on the time delay between the two modes
of GWs are given. In Sec.~\ref{ sec:global } we carry out a global analysis of
the SME coefficients using Monte Carlo methods. The global limits are
simultaneously obtained, breaking the degeneracy among various coefficients.
Section~\ref{ sec:discuss } presents a summary and discusses some possible
future improvements. Throughout the {\it Letter}, we use natural units where
$\hbar = c = 1$.

\section{ Dispersion and birefringence of GWs } 
\label{ sec:disper }

To study the linearized gravity, we first decompose the metric into the
Minkowski metric and a metric perturbation, that is, $g_{\mu\nu} =
\eta_{\mu\nu}+h_{\mu\nu}$. Then the Lagrangian at dimension $d$ in the pure
gravity sector of SME, containing the LI and LIV terms, is expressed as
\citep{Kostelecky:2016kfm},
\begin{equation}
  {\cal L}_{ {\cal K}^{(d)}} = \frac{1}{4} h_{\mu\nu} \hat{\cal K}^{(d)\mu\nu\rho\sigma} h_{\rho\sigma} \,,
\end{equation}
where $\hat{\cal K}^{(d)\mu\nu\rho\sigma} = {\cal K}^{(d)\mu\nu\rho\sigma
\alpha_1 \alpha_2 \cdots \alpha_{d-2}} \partial_{\alpha_1} \partial_{\alpha_2}
\cdots \partial_{\alpha_{d-2}}$ is an operator at mass dimension $d\ge 2$ with
tensor coefficients ${\cal K}^{(d)\mu\nu\rho\sigma \alpha_1 \alpha_2 \cdots
\alpha_{d-2}}$. Obviously, because the indices ``$\mu\nu$'' and ``$\rho\sigma$''
will be contracted with the symmetric metric perturbation $h_{\mu\nu}$, the
two pairs of indices are symmetric. Dedicated investigation
\citep{Kostelecky:2016kfm} showed that there are 14 independent classes of
operators that control the behavior of GWs. However, the operators still need to
obey the infinitesimal gauge transformation $h_{\mu \nu} \rightarrow h_{\mu
\nu}+\partial_{(\mu} \xi_{\nu)}$ where $\xi^{\nu}$ is an arbitrary infinitesimal
vector field. It results in that only 3 of the 14 classes of operators survive.
Summing the Lagrangian at all mass dimensions, \citet{Kostelecky:2016kfm}
obtained the Lagrangian,
\begin{align}
    {\cal L} = {\cal L}_0 &+ \frac{1}{4} h_{\mu\nu} \big( \hat{s}^{\mu\rho\nu\sigma} + \hat{q}^{\mu\rho\nu\sigma} +  \hat{k}^{\mu\nu\rho\sigma} \big) h_{\rho\sigma} \,\label{lagrangian},
\end{align}
where ${\cal L}_0 = \frac{1}{4} \epsilon^{\mu\rho\alpha\kappa}
\epsilon^{\nu\sigma\beta\lambda} \eta_{\kappa\lambda} h_{\mu\nu} \partial_\alpha
\partial_\beta h_{\rho\sigma}$ is the linearized Einstein-Hilbert part.
The remaining three LIV operators have similar forms like $\hat{\cal
K}^{(d)\mu\nu\rho\sigma}$, but the positions of the contracted indices and their
symmetries are different. Their specific forms and properties are discussed in
\citet{Kostelecky:2016kfm}.

With the Lagrangian~\eqref{lagrangian}, it is straightforward to obtain
equations of motion in vacuum \citep{Mewes:2019dhj}. For plane waves, applying
Fourier transformation turns the derivative $\partial_\mu$ into the 4-momentum,
changing these differential equations into algebraic equations of $p^\mu =
(\omega,\boldsymbol{p})$. Assuming that the violation coefficients are small,
\citet{Mewes:2019dhj} found the solution in the temporal gauge and derived the
dispersion relation,
\begin{equation}
    \omega = \left (1-\varsigma^0 \pm \sqrt{|\varsigma_{(+4)}|^2+|\varsigma_{(0)}|^2}\right)\,p\,,
    \label{4-momentum}
\end{equation}
where the leading part gives the ordinary dispersion relation $\omega = p$,
which means that GWs travel at the speed of light. The other terms are
combinations of $\hat{s}^{\mu\rho\nu\sigma}$, $\hat{q}^{\mu\rho\nu\sigma}$, and
$\hat{k}^{\mu\nu\rho\sigma}$, with the derivatives replaced by 4-momentums,
representing the contributions of gLIV.

In the helicity basis, $\varsigma^0$ and $\varsigma_{(0)}$ are real and have
zero helicity, while $\varsigma_{(+4)}$ is complex and has helicity +4
\citep{Mewes:2019dhj}. Taking the ordinary dispersion relation at leading order,
we can decompose the frequency $\omega$ and the propagation direction
$\hat{\boldsymbol{{p}}}$ in the violating terms. Using the spin-weighted
spherical harmonics $_sY_{jm}$ to expand the direction dependent part,
\citet{Mewes:2019dhj} got the explicit series expansion of the violating terms,
\begin{align}
    \varsigma^0 &= \sum_{djm}(-1)^j\,\omega^{d-4}\,_0Y_{jm}(\hat{\boldsymbol{{p}}})\,k^{(d)}_{(I)jm}\,,\notag\\
    \varsigma_{(+4)} &= \sum_{djm}(-1)^j\,\omega^{d-4}\,_{-4}Y_{jm}(\hat{\boldsymbol{{p}}})\,\big(k^{(d)}_{(E)jm}+ik^{(d)}_{(B)jm}\big)\,,\notag\\    
    \varsigma_{(0)} &= \sum_{djm}(-1)^j\,\omega^{d-4}\,_0Y_{jm}(\hat{\boldsymbol{{p}}})\,k^{(d)}_{(V)jm}\,.
\label{SME_coefficients}
\end{align}
Coefficients $k^{(d)}_{jm}$ are the so-called SME coefficients that control the
gLIV. All coefficients are complex numbers and obey $\ k^{(d)}_{j(-m)} =
(-1)^mk^{(d)*}_{jm}$. For the indices, $m$ takes $-j,\cdots,j$, while $j$ takes
$|s|,\cdots, d-2$, with $s$ being the spin weight in $_sY_{jm}$. The  valid
values of the mass dimension $d$, are more complicated: (i) $k^{(d)}_{(I)jm}$
only exists at even dimensions with $d\geq 4$;  (ii) $k^{(d)}_{(V)jm}$ only
exists at odd dimensions with $d\geq 5$; and (iii) $k^{(d)}_{(E)jm}$ and
$k^{(d)}_{(B)jm}$ only exist at even dimensions with $d\geq 6$.

From Eq.~\eqref{4-momentum}, one obtains the phase speed of GWs,
\begin{align}
     v_{\pm} = 1-\varsigma^0\pm|\vec{\varsigma} \, |\,&,\label{speed}
\end{align}
\text{where}\quad $|\vec{\varsigma} \, | \equiv \sqrt{|\varsigma_{(+4)}|^2
+|\varsigma_{(0)}|^2}.$ Like \citet{Shao:2020shv}, we assume that gLIV mainly
occurs at a specific dimension, so it is convenient to define a new parameter
$\varsigma^{(d)}(\hat{\boldsymbol{{p}}}) \equiv
\varsigma(\hat{\boldsymbol{{p}}})\,\omega^{4-d}$, which only depends on the
propagation direction of GWs. Then we can separate the frequency and
direction parts, and Eq.~\eqref{speed} simplifies to
\begin{equation}
    v_{\pm} = 1-\omega^{d-4}\big(\varsigma^{(d)0}\mp|\vec{\varsigma} \, |^{(d)}\big)\,.
\end{equation}
Note that the terms in brackets are only dependent on direction now.

The focus of our study lies in the birefringence phenomenon, occurring when
$|\vec{\varsigma} \, |^{(d)}\neq 0$. This can be divided into two simplified
situations: (i) $\varsigma_{(0)}^{(d)}\neq 0$, $\varsigma_{(+4)}^{(d)}= 0$, and
(ii) $\varsigma_{(+4)}^{(d)}\neq 0$, $\varsigma_{(0)}^{(d)}= 0$. They correspond
to odd dimensions with $d\geq 5$ and even dimensions with $d\geq 6$
respectively. When birefringence occurs, the two modes propagate at different
speeds $v_+\ \text{and}\ v_-$, which will cause an extra time difference
between them after traveling from source to detector. In an expanding universe,
\citet{Kostelecky:2016kfm} derived the time difference between two modes,
\begin{equation}
    \Delta t = 2\,|\vec{\varsigma} \, |^{(d)}\Big(\frac{\omega}{1+z}\Big)^{d-4}\int^z_0\frac{(1+z')^{d-4}}{H(z')}\text{d}z' \,,
    \label{deltat}
\end{equation}
where $|\vec{\varsigma} \, |^{(d)}$ is a function of the SME coefficients,
$\omega$ is the frequency when GWs were emitted, $z$ is the cosmological
redshift, and $H(z)$ is the Hubble parameter. Using the $\Lambda$CDM model, we
have $H(z) = H_0\sqrt{\Omega_{\rm m}(1+z)^3+\Omega_\Lambda}$, where
$H_0\approx 67.4\ \text{km}\,\text{s}^{-1}\,\text{Mpc}^{-1}$ is the Hubble
constant, $\Omega_{\rm m} \approx 0.315$ is the fraction of matter density and
$\Omega_\Lambda \approx 0.685$ is the fraction of vacuum energy density. Once
the SME coefficients and the information of GWs are given, with
Eq.~\eqref{deltat}, we can calculate the time difference between two modes for
every GW event. 

\section{ Constraining time delay of birefringence} 
\label{ sec:costrain }

As discussed in the previous section, a general gLIV will result in anisotropy,
dispersion, and birefringence for GWs, wherein the anisotropic birefringence
phenomenon can be used to constrain the SME coefficients very well.
\citet{Kostelecky:2016kfm} first attempted to impose a restriction on SME
coefficients at $d = 5$ and $6$ using GW150914. At the amplitude peak, they
found no indication of birefringent splitting. Therefore, the time difference
must be less than the width of the peak, which is approximately 3\,ms. They
conservatively estimated that the frequency $f$ was 100\,Hz near the peak.
Substituting these values into Eq.~\eqref{deltat}, they obtained the first set
of constraints. For GWs in GWTC-1, \citet{Shao:2020shv} carried out the first
global analysis of the SME coefficients for GWs and significantly improved the
bounds. Following them, here we use 50 GW events and constrain fully the
coefficient space at mass dimensions $d = 5$ and 6.

For the source parameters of GWs in GWTC-1, we follow the considerations in
\citet{Shao:2020shv} and use the posterior samples provided by the LIGO/Virgo
Collaboration \citep{LIGOScientific:2018mvr}. We use the {\sc
Overall\_posterior} samples  for the 10 binary black holes (BBHs), and the {\sc
IMRPhenomPv2NRT\_lowSpin\_posterior} samples for the binary neutron star (BNS),
GW170817. For events in GWTC-2, we use the {\sc PublicationSamples} dataset in
the posterior files, which were actually used to create results in
\citet{LIGOScientific:2020ibl}. We have checked that, in practice, different
waveforms do not have significant impacts on our final results
\citep{Shao:2020shv}. 

For the frequency at amplitude peak of the chirp signal, we use the fitting
formula in the Appendix A.3 in \citet{Bohe:2016gbl}. However, the formula was
obtained from BBH simulations, which means that when the GW source is a BNS or a
neutron star--black hole (NS-BH) binary, the matter effects of NSs are not taken
into account. Therefore, we choose a lower frequency, $800\,\text{Hz}$, for
GW170817 as it was roughly the sensitivity upper edge~\citep{Shao:2020shv}.  For
other sources whose lighter component mass is less than 3\,$M_\odot$, GW190425
\citep{LIGOScientific:2020aai}, GW190426\_152155, and GW190814
\citep{LIGOScientific:2020zkf}, which may be of BNS or NS-BH origins, we give
rough estimates through the last few cycles near merger of the template
waveforms or the real strain data. We use $800\,\text{Hz}$ for GW190425,
$700\,\text{Hz}$ for GW190426\_152155, and $280\,\text{Hz}$ for GW190814. These
values are rather conservative, because higher frequency will give better limits
according to Eq.~\eqref{deltat}.

For the time difference between two helicity modes, we adopt the same criterion
as in \citet{Shao:2020shv}. For a definite GW event, we assume that the time
delay between two modes satisfies $|\omega\Delta t| \leq2\pi/\rho$, or
equivalently,
\begin{equation}
    \big|\Delta t\big| \leq \frac{1}{\rho f}\,,
\label{eq:inequality}
\end{equation}
where $f$ is the GW frequency at the amplitude peak in the source frame, $\rho$
is the network signal-to-noise ratio (SNR) that can be obtained in
\citet{LIGOScientific:2018mvr, LIGOScientific:2020ibl}. We defer a refined
criterion, possibly with matched filtering, to future work.

\section{ Global analysis and results} 
\label{ sec:global }

For the birefringence phenomenon, we will discuss the two lowest dimensions 5
and 6 separately for nonminimal gravity in the SME. According to Sec.~\ref{
sec:disper }, the indices of these SME coefficients are obtained as follows.
For $d = 5$, one has $j = 0, 1, 2, 3$, thus there are 10 independent
$k^{(5)}_{(V)jm}$.  However, $k^{(5)}_{(V)jm}$ is real only when $m = 0$, so
there are 16 independent components for $k^{(5)}_{(V)jm}$ in total.  For $d =
6$, note that because $|s|\leq j \leq d-2$ and $-j\leq m \leq j$, $j$ only
takes 4. For the same reason as $d = 5$, there are 9 independent components for
$k^{(6)}_{(E)jm}$ and $k^{(6)}_{(B)jm}$ respectively, thus one has 18
independent components in total.

\begin{figure*}[htp]
    \centering
    \includegraphics[width=18cm]{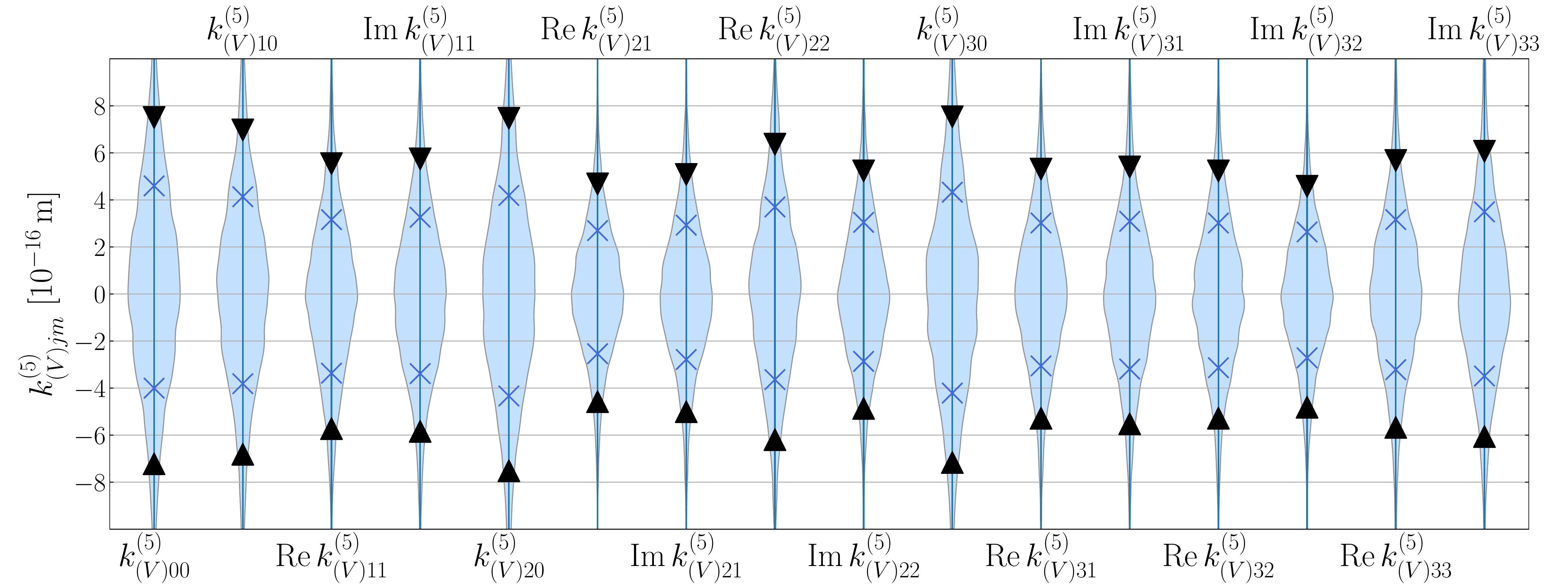}
    \caption{Violin plots of the 16 independent components at mass dimension $d
    = 5$. The 68\% and 90\% confidence intervals are tagged with ``$\times$'' and
    ``$\bigtriangleup$'' respectively.}
    \label{fig:d5}
\end{figure*}

\begin{table}[ht]
\def\arraystretch{1.2}
\caption{The 68\% confidence intervals of the 16 independent components of
$k^{(5)}_{(V)jm}$.}
\centering
\begin{tabular}{cccc }
\hline\hline
$j$ & $m$ &\ \  Component $\ $ & Confidence interval $[10^{-16}\,\text{m}]$ $\ $\\
\hline

    0 & 0 & $k^{(5)}_{(V)00}$ & $(-4.00,\, 4.60)$\\
1 & 0 & $k^{(5)}_{(V)10} $& $(-3.81,\, 4.14)$\\
1 & 1 & $\text{Re}\,k^{(5)}_{(V)11}$ & $(-3.37,\, 3.16)$\\
    &  & $\text{Im}\,k^{(5)}_{(V)11}$ & $(-3.38,\, 3.26)$\\
2 & 0 & $k^{(5)}_{(V)20}$ & $(-4.34,\, 4.19)$\\
2 & 1 & $\text{Re}\,k^{(5)}_{(V)21}$ & $(-2.54,\, 2.70)$\\
    &  & $\text{Im}\,k^{(5)}_{(V)21}$ & $(-2.78,\, 2.93)$\\
2 & 2 & $\text{Re}\,k^{(5)}_{(V)22}$ & $(-3.64,\, 3.70)$\\
    &  & $\text{Im}\,k^{(5)}_{(V)22}$ & $(-2.86,\, 3.05)$\\
3 & 0 & $k^{(5)}_{(V)30}$ & $(-4.21,\, 4.33)$\\
3 & 1 & $\text{Re}\,k^{(5)}_{(V)31}$ & $(-3.06,\, 3.02)$\\
    &  & $\text{Im}\,k^{(5)}_{(V)31}$ & $(-3.19,\, 3.09)$\\
3 & 2 & $\text{Re}\,k^{(5)}_{(V)32}$ & $(-3.13,\, 3.02)$\\
    &  & $\text{Im}\,k^{(5)}_{(V)32}$ & $(-2.71,\, 2.64)$\\
3 & 3 & $\text{Re}\,k^{(5)}_{(V)33}$ & $(-3.22,\, 3.16)$\\
    &  & $\text{Im}\,k^{(5)}_{(V)33}$ & $(-3.49,\, 3.49)$\\
\hline
\end{tabular}\label{table:d5}
\end{table}

\begin{figure*}[htp]
    \centering
    \includegraphics[width=18cm]{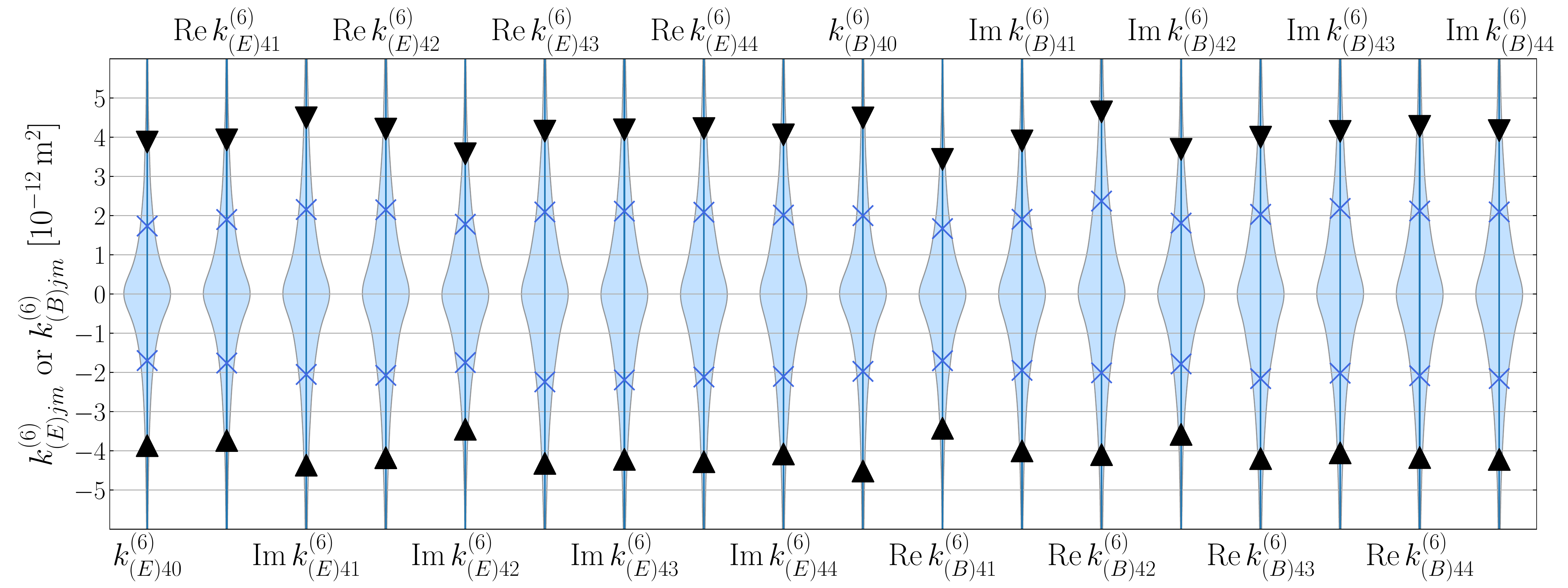}
    \caption{ Same as Fig.~\ref{fig:d5}, for mass dimension $d=6$.  }
    \label{fig:d6}
\end{figure*}

\begin{table}[htbp]
    \def\arraystretch{1.2}
    \caption{ Same as Table~\ref{table:d5}, for $k^{(6)}_{(E)jm}$ and
    $k^{(6)}_{(B)jm}$.  }
    \centering
    \begin{tabular}{cccc}
    \hline\hline
    $j$ & $m$ &\ \  Component $\ $ & Confidence interval $[10^{-12}\,\text{m}^2]$ $\ $\\
    \hline
    4 & 0 & $k^{(6)}_{(E)40}$ & $(-1.70,\, 1.73)$\\
    4 & 1 & $\text{Re}\,k^{(6)}_{(E)41}$ & $(-1.76,\, 1.90)$\\
        &  & $\text{Im}\,k^{(6)}_{(E)41}$ & $(-2.05,\, 2.16)$\\
    4 & 2 & $\text{Re}\,k^{(6)}_{(E)42}$ & $(-2.08,\, 2.15)$\\
        &  & $\text{Im}\,k^{(6)}_{(E)42}$ & $(-1.75,\, 1.78)$\\
    4 & 3 & $\text{Re}\,k^{(6)}_{(E)43}$ & $(-2.24,\, 2.09)$\\
        &  & $\text{Im}\,k^{(6)}_{(E)43}$ & $(-2.19,\, 2.11)$\\
    4 & 4 & $\text{Re}\,k^{(6)}_{(E)44}$ & $(-2.12,\, 2.09)$\\
        &  & $\text{Im}\,k^{(6)}_{(E)44}$ & $(-2.10,\, 2.01)$\\
    4 & 0 & $k^{(6)}_{(B)40}$ & $(-1.97,\, 2.00)$\\
    4 & 1 & $\text{Re}\,k^{(6)}_{(B)41}$ & $(-1.70,\, 1.67)$\\
        &  & $\text{Im}\,k^{(6)}_{(B)41}$ & $(-1.95,\, 1.91)$\\
    4 & 2 & $\text{Re}\,k^{(6)}_{(B)42}$ & $(-2.01,\, 2.37)$\\
        &  & $\text{Im}\,k^{(6)}_{(B)42}$ & $(-1.78,\, 1.81)$\\
    4 & 3 & $\text{Re}\,k^{(6)}_{(B)43}$ & $(-2.16,\, 2.03)$\\
        &  & $\text{Im}\,k^{(6)}_{(B)43}$ & $(-2.02,\, 2.18)$\\
    4 & 4 & $\text{Re}\,k^{(6)}_{(B)44}$ & $(-2.09,\, 2.12)$\\
        &  & $\text{Im}\,k^{(6)}_{(B)44}$ & $(-2.16,\, 2.09)$\\
    \hline
    \end{tabular}\label{table:d6}
    \end{table}

First, we take $d = 5$ to show how 50 GW events can constrain the coefficient
space. For every GW event, based on Eqs.~(\ref{deltat}--\ref{eq:inequality}) we
have an inequality of the 16 components, and each inequality gives an area
between two hypersurfaces symmetrical about the origin.\footnote{For $d = 5$,
they are hyperplanes, while for $d=6$ they are more complicated.} It is easy to
see that $\Delta t$ is actually a linear combination of $k^{(5)}_{(V)jm}$, and
the combination coefficients are the spin-weighted spherical harmonics,
$_0Y_{jm}$, which are direction-dependent. Based on the fact that GW events
scatter in different sky areas, these linear combinations are expected to be
linearly independent.  Consequently, the 50 pairs of hypersurfaces can enclose a
region containing the origin, where all SME coefficients must be. In this way we
have completed a simple but global limitation of the coefficient space. In this
closed area every coefficient is\textit{ bounded}. Then we can obtain the value
range of each parameter in principle.

To incorporate all uncertainties and statistically obtain the distribution of
these SME coefficients, one requires the distribution of the ``measured'' time
delay. As a conservative consideration, we can take the upper bound of $\Delta
t$ in Eq.~\eqref{deltat} as its standard deviation. Because the violations in
SME are expected to be tiny, we take the central value of $\Delta t$ to be zero.
We assume that the ``measured'' time delay obeys a Gaussian normal distribution
with $\mu = 0$ and $\sigma = {1}/({\rho f})$. Based on this, we carry out a
global analysis following the steps below.
\begin{enumerate}[(i)]
    \item For each GW event, we randomly draw the source parameters in the
    posterior samples, and calculate the frequency at the amplitude peak.
    \item We randomly draw the time delays of the 50 GW events from their
    respective distributions and take these time delays as the result of one
    measurement in the Monte Carlo processes.
    \item Given the SME coefficients, we calculate the theoretical time delay
    $\Delta t^{(\text{th})}_{n}$ for the $n$-th GW event where $n$ takes $1,
    2,\cdots,50$. 
    \item Considering that the measured time delay obeys a Gaussian distribution
    with $\mu = \Delta t^{(\text{th})}_n$ and $\sigma ={1}/({{\rho_n} {f_n}})$,
    we construct the multi-Gaussian likelihood as a function of the SME
    coefficients for all events. After maximizing the likelihood, the
    corresponding SME coefficients at the extreme point are what we extract from
    each draw.
    \item We repeat steps (i)--(iv) enough times until a stable distribution is
    obtained.
\end{enumerate}

For the 16 components at $d = 5$, the violin plots of the one-dimensional
marginalized distributions are given in Fig.~\ref{fig:d5}, while their
confidence intervals are given in Table~\ref{table:d5}. Because the dependence
of $\Delta t$ on the SME coefficients is linear, the marginalized distributions
are very close to Gaussian distributions. Similar to \citet{Shao:2020shv}, no
obvious correlation is found among the components due to the different
properties of the GW events, including  sky locations, peak GW frequencies and
so on.

For the 18 components at mass dimension $d = 6$, we have to analyze with more
care. Since $\varsigma_{(+4)}$ is a complex number, the calculation of
$\sqrt{|\varsigma_{(+4)}|^2}$ is much more difficult than $d = 5$, whereas
$\varsigma_{(0)}$ is real and the square root is trivial. Due to this reason,
the dependence of $\Delta t$ on the SME coefficients is highly nonlinear,
leading the marginalized distributions to be more peaked than the Gaussian
distribution in the end. Despite the complications above, correlations between
the 18 independent components are still small. Their one-dimensional
marginalized distributions are given in Fig.~\ref{fig:d6} and the 68\%
confidence intervals are illustrated in Table~\ref{table:d6}.

\section{ Discussion and summary }
\label{ sec:discuss }

Gravitational waves (GWs) provide excellent opportunities to make a profound
study of the gravitational Lorentz invariance violation (gLIV). Lots of work
\citep{Wang:2020pgu, Nishizawa:2017nef,  LIGOScientific:2019fpa,
Mirshekari:2011yq, Wang:2020cub} have been performed to test the Lorentz
symmetry in General Relativity (GR) using GW events. While  a general gLIV
includes anisotropy, dispersion, and birefringence, previous work mostly
emphasized the isotropic birefringent phenomenon.

In our work, we focus on the anisotropic birefringence of GWs in the popular
theoretical framework called the standard-model extension
\citep[SME;][]{Colladay:1996iz, Kostelecky:2003fs}. Assuming that gLIV mainly
occurs at a particular mass dimension, for the two lowest dimensions that can
produce birefringence, $d = 5$ and 6, we separately carry out global analysis on
them and constrain the whole coefficient space fully. Thanks to the multiple GW
events in the GW transient catalogs GWTC-1 and GWTC-2, we completely break the
degeneracy among the 16 independent components of the SME coefficients
$k^{(5)}_{(V)jm}$ at $d = 5$, and 18 independent components at $d = 6$.  This is
reflected in the results of Tables~\ref{table:d5}~\&~\ref{table:d6}, which are
obtained by \textit{simultaneously} constraining all the coefficients, instead
of setting only one coefficient to be nonzero each time \citep[i.e., the
``maximum reach'' approach; ][]{Tasson:2019kuw}. As can be seen, there is no
obvious violation of GR in our results, and our limits on the SME coefficients
are $2$ to $7$ times tighter than that in the earlier work \citep{Shao:2020shv}.

While in this paper, we have only made use of one type of phenomena in the
birefringent gLIV, namely the differences of the arrival times between two
modes, there are many more approaches to improve our tests. First, more
information can be extracted from the birefringence phenomena, such as the
changes of the polarizations. Similar analysis has been conducted in the photon
sector \citep{Kostelecky:2008be, Kostelecky:2013rv}. Second, Eq.~\eqref{deltat}
is a relatively simple estimation of the time delay~\citep{Shao:2020shv}, and in
fact \citet{Mewes:2019dhj} has worked out the whole deformed waveform analytics
and we can use these modified templates to implement the matched-filtering
analysis. 
For mass dimensions $5$ and $6$, the
computational cost is expected to be significant \citep{ONeal-Ault:2021uwu}.
Third, in a very unlikely scenario, the pattern function of GW detectors
might hide one of the two polarizations. However, we think it is impossible to
happen in all 50 GW events that we used here. However, to entirely rule out this
possibility, one needs to incorporate the polarization of the events, probably
in a matched-filtering analysis~\citep{Wang:2020cub, ONeal-Ault:2021uwu}. 
Finally, advanced detectors
like the KAGRA \citep{Aso:2013eba} have recently joined the global efforts to
detect GWs. With the ground-based detector network, more and more GWs in a wider
frequency range will obviously bring direct improvements to our results. 

In conclusion, the GWs are undoubtedly fantastic phenomena to test fundamental
principles in modern physics, and we expect a brighter future in the coming
years for the fundamental physics through the use of these treasures of the
Nature.

\begin{acknowledgments}
We are grateful to Yi-Fan Wang for discussions and 
the LIGO/Virgo Collaboration for providing the
posterior samples of their parameter-estimation studies. This work was supported
by the National Natural Science Foundation of China (11975027, 11991053,
11721303), the National SKA Program of China (2020SKA0120300), the Young Elite
Scientists Sponsorship Program by the China Association for Science and
Technology (2018QNRC001), the Max Planck Partner Group Program funded by the Max
Planck Society, and the High-performance Computing Platform of Peking
University. ZW is supported by the Principal's Fund for the Undergraduate
Student Research Study at Peking University.
\end{acknowledgments}

\bibliography{refs}{}
\bibliographystyle{aasjournal}

\end{document}